\pdfoutput=1

\newcommand{\degr}{$^\circ$}
\newcommand{\caproman}[1]{\uppercase\expandafter{\romannumeral#1}}

\newcommand{\BGmail}{bernardgottschalk\,@\,gmail.com}

\newcommand{\eqn}[1]{Eq.\,(\ref{eqn:#1})}
\newcommand{\eqns}[2]{Eqs.\,(\ref{eqn:#1}-\ref{eqn:#2})}
\newcommand{\fig}[1]{Fig.\,\ref{fig:#1}}

\documentclass{article}

\title{\bf Global surface temperature trends\\and the effect of World War \caproman{2}:\\a parametric analysis\\(long version)}
\author{Bernard Gottschalk\thanks{Harvard University Laboratory for Particle Physics and Cosmology, 18 Hammond St., Cambridge, MA 02138, USA, \BGmail}}
\date{\today}

\usepackage{amsmath}
\usepackage{latexsym}
\usepackage{graphicx}

\graphicspath{%
    {converted_graphics/}
    {/}
}
\begin{document}

\maketitle
\vspace{.15in}
\begin{center}{\large\bf for Haim Goldberg (1939$-$2017)}\end{center}
\vspace{.30in}

\begin{abstract}
\noindent We fit twelve independent global surface temperature monthly time series with a Gaussian on a quadratic background. The data, published by NOAA, comprise land based (atmospheric) series from 1880 through 2016 in six contiguous 30\degr \ zones of latitude north to south, and ocean series in the same six zones. 

The nonlinear least-squares fits use identical fitting technique (including initial values) for all zones. The fits yield six parameters, with errors, which describe the bump and the underlying trend in global surface temperature. The polar zones have inadequate data. The remaining eight each show evidence of a statistically significant bump coincident with World War \caproman{2}.

The final fit excluding four polar zones (insufficient data) and one other (atypical fit) is
\[T(t)\;=\;p_1+p_2\,u+p_3\,u^2+p_6\;e^{\textstyle{-0.5((u-p_7)/p_8)^2}}\]
where $T$ is temperature (\degr C), $t$ is time (yr), $u$ is a dimensionless transformed time
\[u\;\equiv\;B\,(t-A)\]
\[A\;=\;1948.5\;\mathrm{yr}\quad,\quad B\;=\;0.014607\;{\mathrm{yr}}^{-1}\]
and
\[p_1\;=\;-0.3752\pm0.0157\;^\circ\mathrm{C}\quad,\quad p_2\;=\;0.5112\pm0.0162\;^\circ\mathrm{C}
  \quad,\quad p_3\;=\;0.4210\pm0.0334\;^\circ\mathrm{C}\]
\[p_6\;=\;0.3394\pm0.0652\;^\circ\mathrm{C}\quad,\quad p_7\;=\;-0.0692\pm0.0062
  \quad,\quad p_8\;=\;0.0276\pm0.0064\]
(1$\sigma$ errors). Thus the Gaussian amplitude differs from zero by 5.2 standard deviations. The area under the Gaussian is $1.948\pm0.577$\;\degr C\,yr differing from zero by 3.4 standard deviations.
The fit extrapolates to a 0.5\;\degr C rise over the next 20\;years if conditions remain the same.

\end{abstract}

\clearpage
\section{Introduction}
Time series of global surface temperature often exhibit a bump coincident with World War \caproman{2} (WW2), as came to our attention on the first page of the 1/19/17 New York Times. Taken at face value, this is direct evidence of the effect of human activity on global temperature. There is, of course, ample indirect evidence.

In this paper we show that the bump is a robust feature of the surface temperature record. It shows up in eight independent time series, four land-based and four ocean-based. We use curve fitting (parametric analysis), fitting a Gaussian on a quadratic background independently to each data set. That provides independent error estimates for each of the six fit parameters and for each data set. 

\section{Method}

\subsection{Data}
We selected twelve files from a NOAA website \cite{ERSSTv4}, half land-based and half ocean (ERSSTv4), each covering the globe in six latitude zones of width 30\degr . Table \ref{tbl:files} lists the file names and the index we assigned to each, grouping them as land N-S (1-6) and ocean N-S (7-12). Each file is statistically independent of the others, and land/ocean separation allows us to look for corresponding effects separately and avoids any error that might result from combining land with ocean. Our overall approach is to fit each file separately and, only afterwards, find the weighted mean of the seven of twelve fits we decide to accept. (Data for the four polar sets 1, 6, 7, 12 are simply too poor and we exclude the land-based equatorial zone 3 for reasons discussed later.)  

\subsection{Parameterization}
Let $t_i$ and $T_i$ be, respectively, the time of the $i^\mathrm{th}$ measurement in years and the temperature anomaly in \degr C. The time range 1880$-$2017 is awkward for a polynomial fit because the coefficients will vary by orders of magnitude ($2017^2$ is a large number) and have rather obscure meanings. The first coefficient, for instance, is the temperature at year 0. 

Therefore, rather than fit $T_i(t_i)$ directly, we transform $t$ to a dimensionless scaled variable $u$ ranging from -1 to 1 as $t$ ranges from 1880 to 2017. Because 1 to any power is 1, the resulting polynomial coefficients are all of the same order of magnitude and much easier to interpret and plot.\footnote{~A more important consequence of the transformation, probably not needed here, is that it greatly improves the precision of high-order polynomial fits because the matrix involved is better conditioned for inversion \cite{nr}.} For instance, the first coefficient is now simply the temperature at midrange, and $T_M$ at the end of range ($t=t_M, u=1$) is simply the sum of the coefficients. 

The time transformation and its inverse are
\begin{equation}\label{eqn:u}
u\;=\;B(t-A)\quad,\quad t=A+(u/B)
\end{equation}
where in our case
\begin{equation}\label{eqn:AB}
 A\equiv(t_M+t_1)/2=1948.5\;\hbox{yr}\quad,\quad B\equiv 2/(t_M-t_1)=0.014607\;\hbox{yr$^{-1}$}
\end{equation}

Let $\vec{p}$ be a set of ten fit parameters. The polynomial is
\begin{equation}\label{eqn:P}
P(u)\;\equiv\;p_1+p_2u+p_3u^2+\ldots\;=\;\sum_{j=1}^N\,p_j\,u^{\,j-1}
\end{equation}
$N$ is the number of terms. We reserve the first five parameters for polynomial coefficients, but will find that only three (quadratic polynomial) are necessary and justified.

The Gaussian is
\begin{equation}\label{eqn:G}
G(u)\;\equiv\;p_6\;e^{\displaystyle{-.5\left((u-p_7)/p_8\right)^2}}
\end{equation}
with $p_6$, $p_7$ and $p_8$ the amplitude, mean and $\sigma$ respectively.

\begin{table}[t]
\setlength{\tabcolsep}{10pt}
\begin{center}
\begin{tabular}{rlr}
\multicolumn{1}{c}{index}&           
\multicolumn{1}{c}{file name}&           
\multicolumn{1}{c}{comment}\\           
\noalign{\vspace{6pt}}
1&   aravg.mon.land.60N.90N.v4.0.1.201612.txt&6 30\degr \ zones N-S\\
2&   aravg.mon.land.30N.60N.v4.0.1.201612.txt\\
3&   aravg.mon.land.00N.30N.v4.0.1.201612.txt\\
4&   aravg.mon.land.30S.00N.v4.0.1.201612.txt\\
5&   aravg.mon.land.60S.30S.v4.0.1.201612.txt\\
6&   aravg.mon.land.90S.60S.v4.0.1.201612.txt\\
\noalign{\vspace{6pt}}
7&   aravg.mon.ocean.60N.90N.v4.0.1.201612.txt&6 30\degr \ zones N-S\\
8&   aravg.mon.ocean.30N.60N.v4.0.1.201612.txt\\
9&   aravg.mon.ocean.00N.30N.v4.0.1.201612.txt\\
10&  aravg.mon.ocean.30S.00N.v4.0.1.201612.txt\\
11&  aravg.mon.ocean.60S.30S.v4.0.1.201612.txt\\
12&  aravg.mon.ocean.90S.60S.v4.0.1.201612.txt
\end{tabular}
\caption{Files used in this study.}\label{tbl:files}
\end{center}
\end{table}

The Gaussian can be replaced by a 5-parameter pulse
\begin{eqnarray}
H(u)\;&\equiv&\;0\hspace{165pt}-\infty< u<p_7\\
&\equiv&p_6(1-e^{\displaystyle{-(u-p_7)/p_9}})\hspace{93pt}0\le u<u_c\\ 
&\equiv&p_6(1-e^{\displaystyle{-(u_c-p_7)/p_9}})\;e^{\displaystyle{-(u-u_c)/p_{10}}\hspace{16pt} u_c\le u<\infty}\\
\noalign{\noindent{where}}
u_c&\equiv&p_7+p_8
\end{eqnarray}
The pulse starts at $p_7$ and rises exponentially towards an asymptote $p_6$ with a time constant $p_9$ for a time $p_8$\,, whereupon it falls exponentially back to the baseline with a time constant $p_{10}$. Five turns out to be more parameters than the data will support, so we eventually fixed the rise and fall time constants $p_9$ and $p_{10}$ at the equivalent of 1 year. Then the pulse is nearly square with amplitude $p_6$, leading corner $p_7$ and trailing corner $p_8$. As long as they are small, the exact choice of $p_9$ and $p_{10}$ hardly affects the fit, equivalent to saying they are not well determined by the data.

We will use the pulse only as one more demonstration of fit robustness. All our final results pertain to the Gaussian on a quadratic background.

\subsection{Curve Fitting Technique}
We use conventional techniques to minimize
\begin{equation}\label{eqn:chisq}
\chi^2\;\equiv\;\sum_{i=1}^M\;(T_i-T(u_i;\vec{p}))^2/\sigma_i^2\;=\;
  \sum_{i=1}^M\;w_i\,(T_i-T(u_i;\vec{p}))^2
\end{equation}
with respect to the set of $n$ parameters $\vec{p}$. The weight $w_i$ of the $i^\mathrm{th}$ point is defined as the inverse of its variance $\sigma_i^2$. Once the best parameters have been found, if
\begin{equation}\label{eqn:csqpdf}
\chi^2/(M-n)\approx1
\end{equation}
the fit is considered good, since \eqn{csqpdf} simply means that the typical distance of each datum from the fit is comparable to the typical error. If $\chi^2/(M-n)\gg1$ the fit is poor. If $\chi^2/(M-n)\ll1$ the fit is `too good'. There are too many parameters and the fit chases random fluctuations in the data points. 

In the present case, \eqn{csqpdf} is too much to hope for, since it depends on an accurate estimate of the variances, exceedingly difficult for temperature anomalies. It also depends on the absence of any systematic (non random) effects not included in the fit function. The best we will achieve is $\chi^2/(M-n)\approx4$. The main systematic effect in that case (zones 9 and 10) is the El Ni\~no Southern Oscillation (ENSO).

We use the standard non-linear least-squares Levenberg-Marquardt algorithm as adapted from Numerical Recipes \cite{nr}. That version allows selected parameters to be held fixed (as desired) at their initial values. It also returns control to the calling program after each pass, whereupon the calling program can either initiate another pass or decide that minimization has converged (in our case, that $\chi^2$ has improved by less than 0.1\%). Then, following Numerical Recipes, the program calls Marquardt once more with $\lambda\doteq0$, after which the variance in each fitted parameter equals the corresponding diagonal term of the curvature matrix.

Non-linear least-squares routines require an initial value for each parameter, that is, the routine must be told where to start looking. In our implementation the initial Gaussian amplitude, location and width are supplied by an input text file, which also furnishes other quantities needed by the program (Fig.\,\ref{fig:inputFile}). Initial polynomial coefficients, by contrast, are found by omitting the Gaussian parameters and performing a preliminary linear least-squares fit, which does not itself require initial values.

\begin{figure}[t] 
\begin{verbatim}
     1  GAUSSIAN   1   10/        GAUSSIAN,PULSE,POLY; wt(0-4); sel trial
     2  0 1 0 2*1 2*0 4*1 0/      skip/use this data file
     3  24    1/                  smoothing sigma (months), # passes
     4  10   20   .001  3*.001/   lambda0, max pass, conv, poly deltas
     5  .9  -.07   .03/           Gaussian ampl,mean,sigma
     6  .001  .001  .001/         Gaussian deltas (3)
     7  .3 -.1  .1  2*.015 /      pulse ampl,t0,delt,tau1,tau2
     8  3*.001  0  0   /          pulse deltas (5)
\end{verbatim}
  \caption{Input file.}\label{fig:inputFile}
\end{figure}

\subsection{Statistics}
For completeness we include a few standard formulas used in the analysis. 

\subsubsection{Moments of a Frequency Distribution}
We will be computing frequency distributions of {\em residuals} namely 
\begin{equation}
r_i\;\equiv\;T_i-T(u_i;\vec{p})
\end{equation}
$r_i$ being the deviation of the $i^\mathrm{th}$ data point from the fit. We first define (say) 100 equal bins of temperature such that the central value $x_1$ of the first bin is the smallest residual and $x_{100}$ is the largest. If, after sorting the residuals into those bins, the $i^\mathrm{th}$ bin contains $N_i$ counts, the first five central moments of the frequency distribution $N(x)$ are, as conventionally defined \cite{nr},
\begin{eqnarray}
\hbox{sum}\;\equiv\;S&\equiv&\sum_{i=1}^{100}N_i\qquad\mathrm{(dimensionless)}\label{eqn:sum}\\
\hbox{mean}\;\equiv\;m&\equiv&\frac{1}{S}\;\sum_{i=1}^{100}N_i\,x_i\qquad(^\circ\mathrm{C})\\
\hbox{rms deviation}\;\equiv\;\sigma_x&\equiv&\left(\frac{1}{S}\;\sum_{i=1}^{100}N_i\,(x_i-m)^2\right)^{1/2}\qquad(^\circ\mathrm{C})\\
\hbox{skewness}\;\equiv\;s&\equiv&\frac{1}{S\;\sigma_x^3}\;\sum_{i=1}^{100}N_i\,(x_i-m)^3\qquad\mathrm{(dimensionless)}\\
\hbox{kurtosis}\;\equiv\;k&\equiv&\frac{1}{S\;\sigma_x^4}\;\sum_{i=1}^{100}N_i\,(x_i-m)^4\;-\;3\qquad\mathrm{(dimensionless)} \label{eqn:kurtosis}
\end{eqnarray}
The last two serve as quantitative tests of the randomness of the $r_i$. If the residuals are random, their frequency distribution should be Gaussian, with zero skewness and kurtosis.

\subsubsection{Weighted Average}
The weighted average of a set of numbers $x_i$ each having an error $\sigma_i$ is \cite{bevington}
\begin{eqnarray}
<x>&=&\textstyle{\sum}\, w_i\,x_i/\sum w_i\\
\noalign{\noindent and its error is}
\sigma_{<x>}&=&(1/\;\textstyle{\sum}\, w_i)^{1/2}\\
\noalign{\noindent where, as always}
w_i&\equiv&1/\sigma_i^2
\end{eqnarray}

\subsection{Computer Program}
The Fortran program (Intel Parallel Studio XE 2016 Update 4) begins by reading an input text file containing initial parameters and other numbers governing the fit (Fig.\,\ref{fig:inputFile}). It then reads and fits (in about one second) all twelve data files, using the same input file for each to avoid bias. The input file is dumped onto an output file which records the fit progress (parameter values after each pass) as well as the weighted average of the final parameters for selected data files. The program also prepares working graphics, so that the run can be assessed immediately, as well as text files for subsequent use by a graphics program and by \LaTeX. 

We have deliberately chosen certain constants to force the program to look fairly hard for the $\chi^2$ minimum. Thus we set the convergence criterion at $0.1\%$ and the initial amplitude at about $3\times$ the final value. The program typically takes $4-6$ passes to converge, except zone\,3 which takes 12 passes to finally arrive at the wrong answer, as we will see.

\section{Results}

\subsection{Survey of the Data}
\begin{figure}[p] 
  \centering
  \includegraphics[bb=0 0 640 480,width=5.32in,height=4.0in,keepaspectratio]{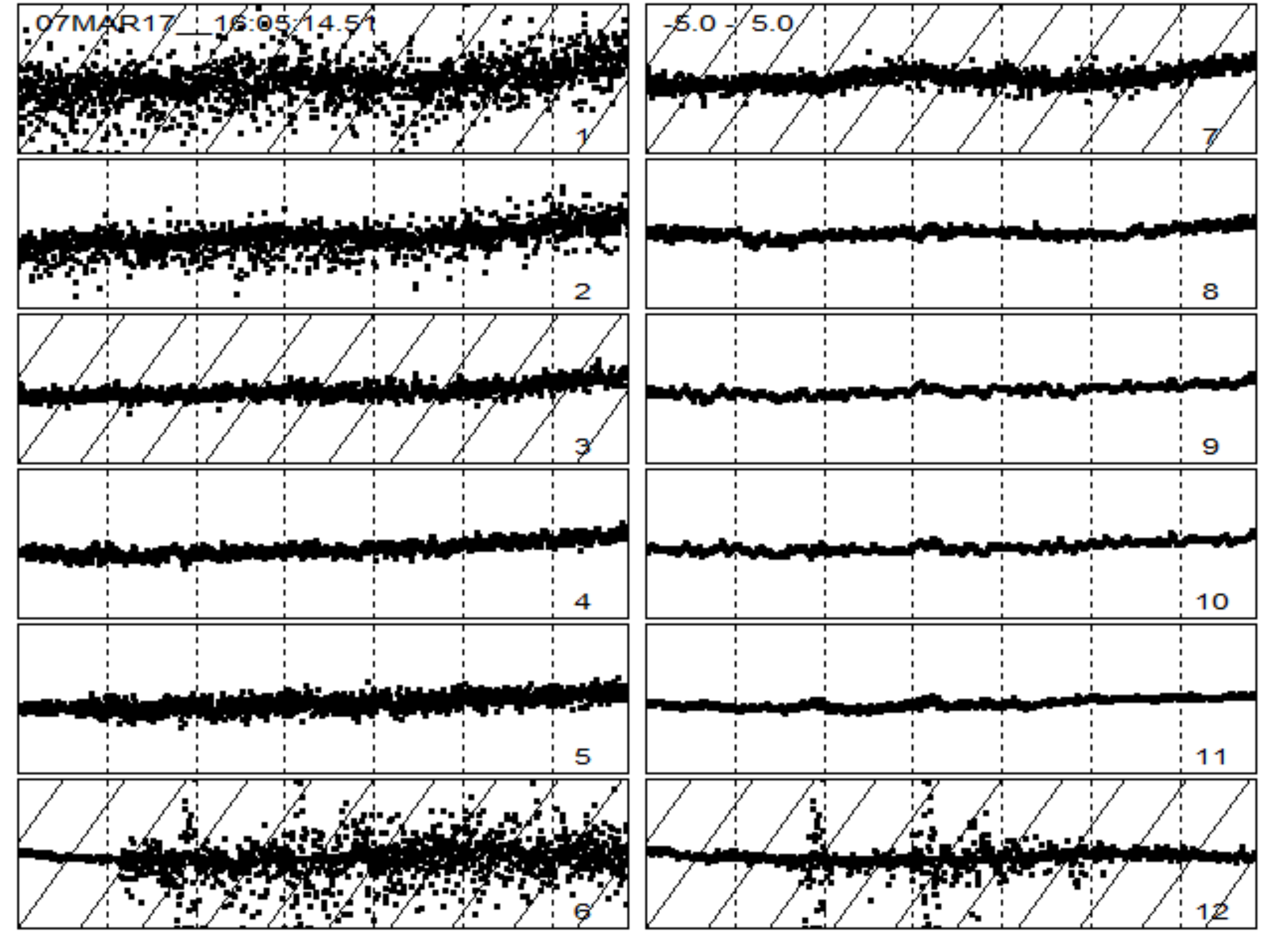}
  \caption{Raw data. Left column: land data, N-S going downwards. Right column: ocean data. Each window is $\pm5$\,\degr C. Cross-hatched sets are excluded from the final cut.}
  \label{fig:multiyy}
\end{figure}

\fig{multiyy} shows the raw data in a layout we will use repeatedly. The right-hand column displays land-based (atmospheric) monthly measurements N$-$S and the left-hand column, ocean-based also N$-$S. All plots in a given figure are to the same ordinate scale, in this case $\pm5$\,\degr C. The abscissa, marked at two-decade intervals, always runs from 1880 through 2016. Latitude zones dropped from the final cut are cross-hatched. As \fig{multiyy} shows, the four polar zones 1,6,7,12 exhibit much more scatter than the others, and are not useful. The reason for dropping land-based zone 3 is explained below. The equatorial ocean-based zones 9,10 have the cleanest data by far. 

\begin{figure}[p] 
  \centering
  \includegraphics[bb=0 0 640 480,width=5.32in,height=4.0in,keepaspectratio]{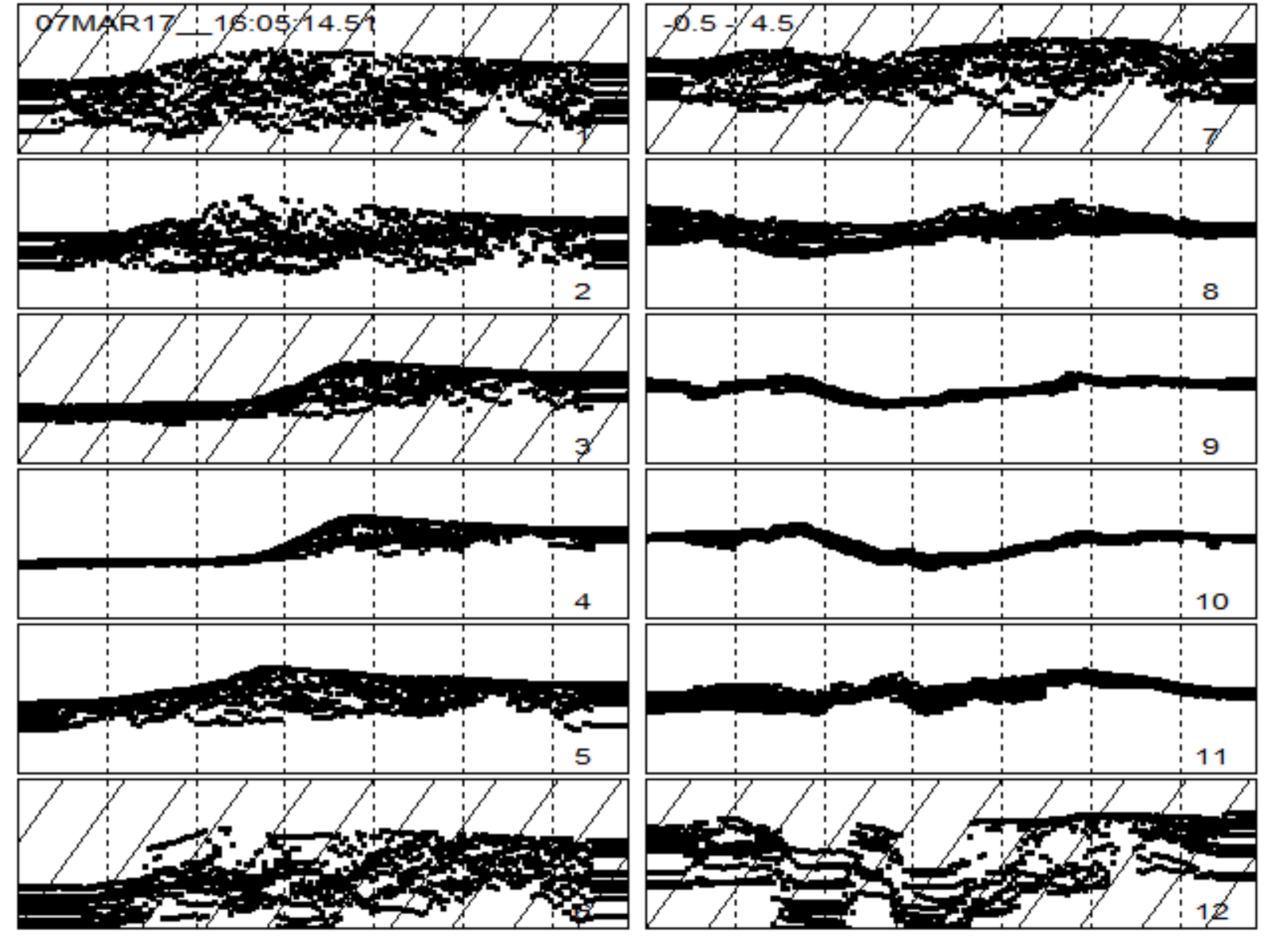}
  \caption{Log$_{10}$ of weights\;=\;1/variance, each window from -0.5 to 4.5 (5 orders of magnitude).}
  \label{fig:multiwl}
\end{figure}

\fig{multiwl} shows the weights (inverse variance) on a semi-logarithmic plot ($\log_{10}=-0.5\;\mathrm{to}\;4.5$). The data files give the overall variance of each point as well as three separate components. We have plotted and used weights based on the overall variance. The range of weights for some zones, even month-to-month, is as large as three orders of magnitude. Again, 9,10 stand out: the variation in weights (about one order of magnitude) is mostly long-term. Fortunately, we will find the fits are rather insensitive to weights. 

\begin{figure}[p] 
  \centering
  \includegraphics[bb=0 0 640 480,width=5.32in,height=4.0in,keepaspectratio]{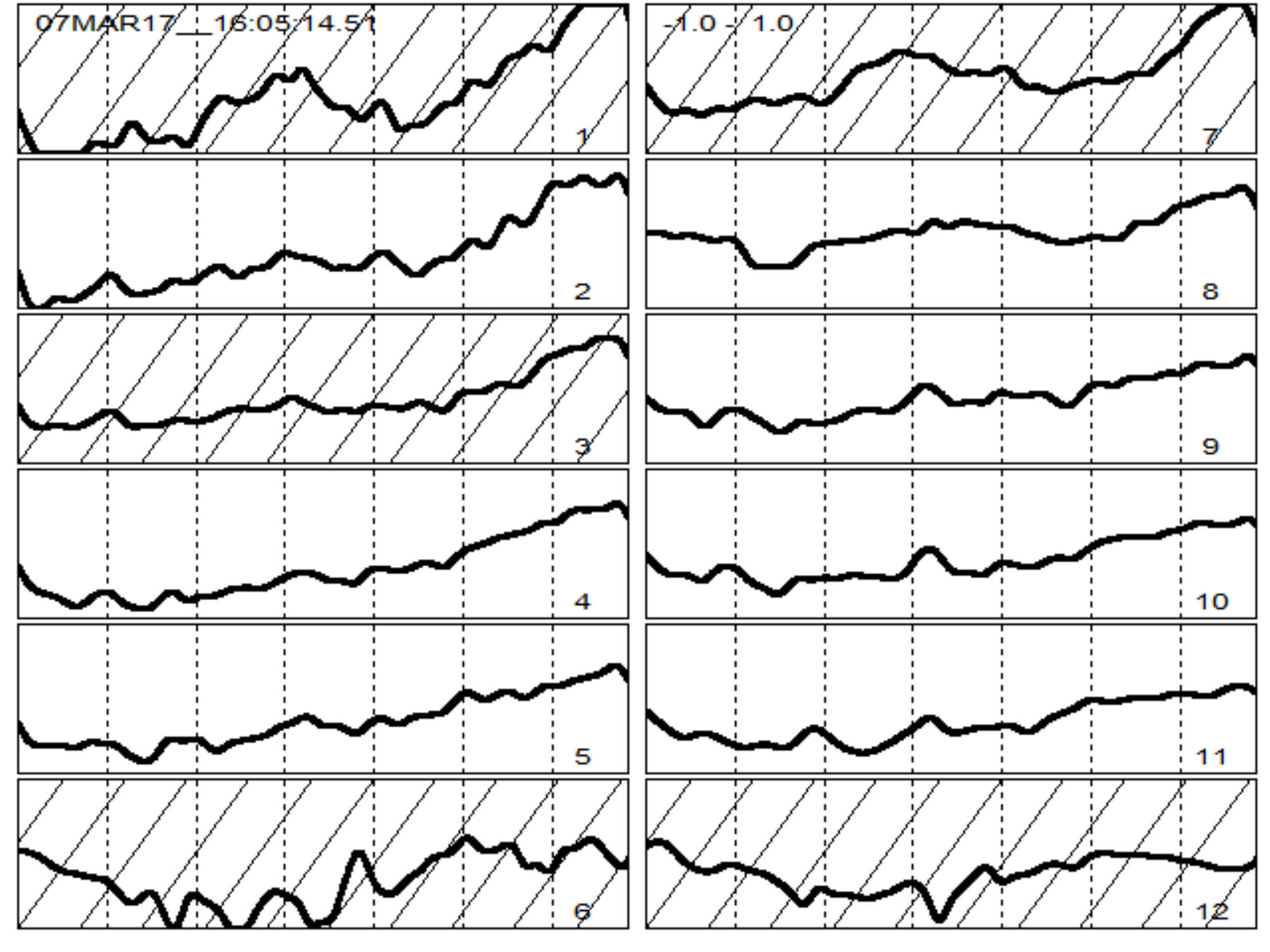}
  \caption{Data smoothed by convolution with a Gaussian of $\sigma=24$\,months.}
  \label{fig:multiys}
\end{figure}

\fig{multiys} shows the raw data (range $\pm1$\,\degr C) as smoothed by convolution with a Gaussian of $\sigma=24$\;months, chosen on the basis of preliminary studies. Zones 9,10 are quite similar and show the WW2 bump clearly. There is a hint of the bump in some other zones. Smoothing can guide us where to look but, by itself, yields no numbers let alone errors.  

Anticipating things somewhat, \fig{multiyf} shows the final fits using input parameters given in \fig{inputFile}. All six parameters (3 for the background and 3 for the Gaussian) are free to change. Zone 3 finds it profitable to spend the extra degrees of freedom on the background rather than the bump. (Even so, the bump is there, as we will see shortly.) We were not able to find a single set of input parameters that avoids this behavior in one or another zone. Therefore our final choice yields useful parameters and errors for only seven zones, and zone 3 is excluded from the global weighted average.

Our preoccupation with WW2 does not rule out a persistent bump somewhere else, which our program would not find on its own. However, a quick search guided by \fig{multiys} turned up nothing.

\subsection{Tests of Robustness}
In \fig{multiyflock} we lock the Gaussian mean and $\sigma$. In other words, the program looks in every zone for a Gaussian of a given width at a given time (suggested, of course, by previous results) but is free to adjust the amplitude and the background polynomial. That yields a zone 3 amplitude and polynomial in fair agreement with the other land-based zones.

\begin{figure}[p] 
  \centering
  \includegraphics[bb=0 0 640 480,width=5.32in,height=4.0in,keepaspectratio]{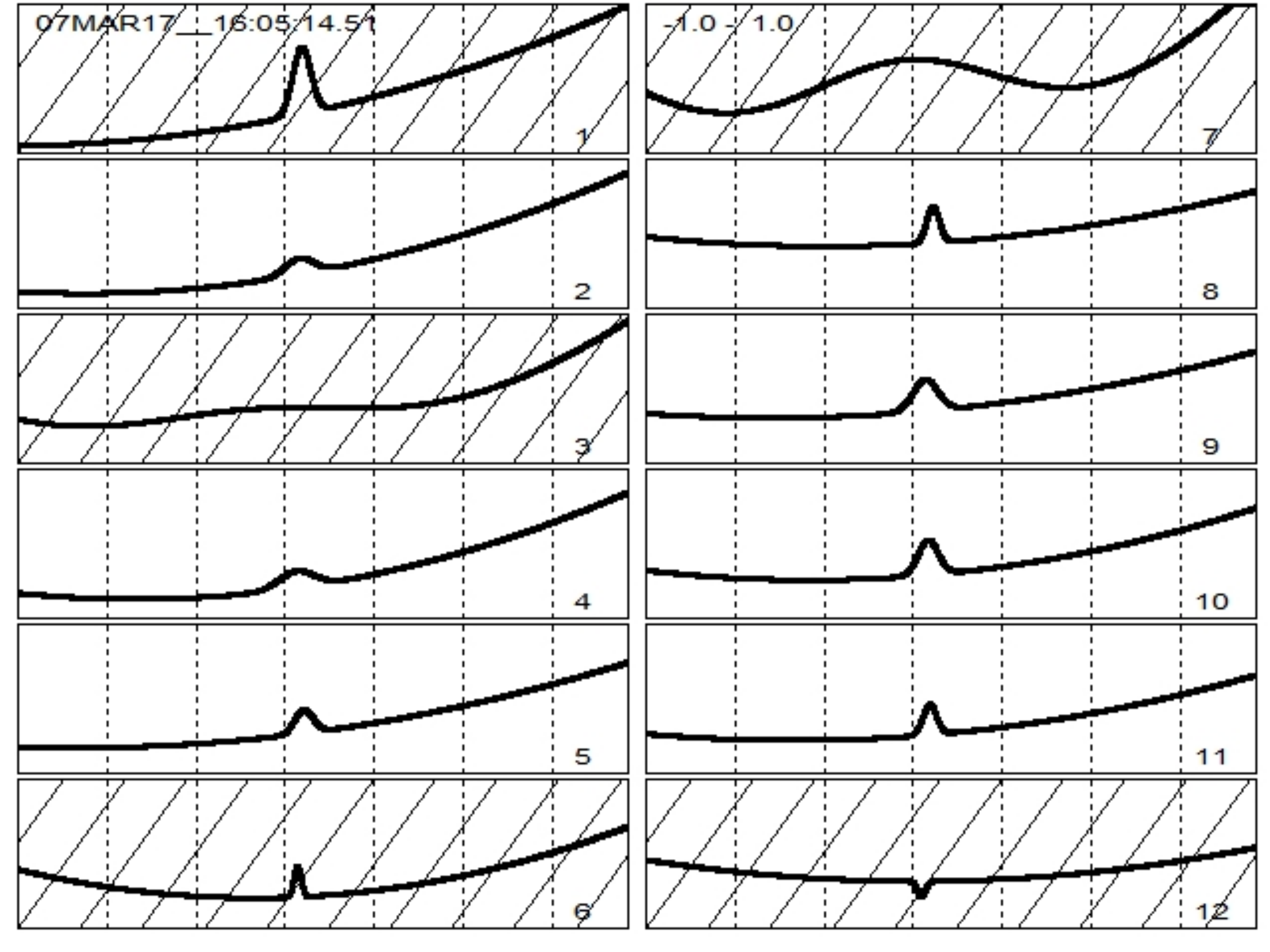}
  \caption{Our final cut: fits with a Gaussian on a quadratic background.}
  \label{fig:multiyf}
\end{figure}

\begin{figure}[p] 
  \centering
  \includegraphics[bb=0 0 640 480,width=5.32in,height=4.0in,keepaspectratio]{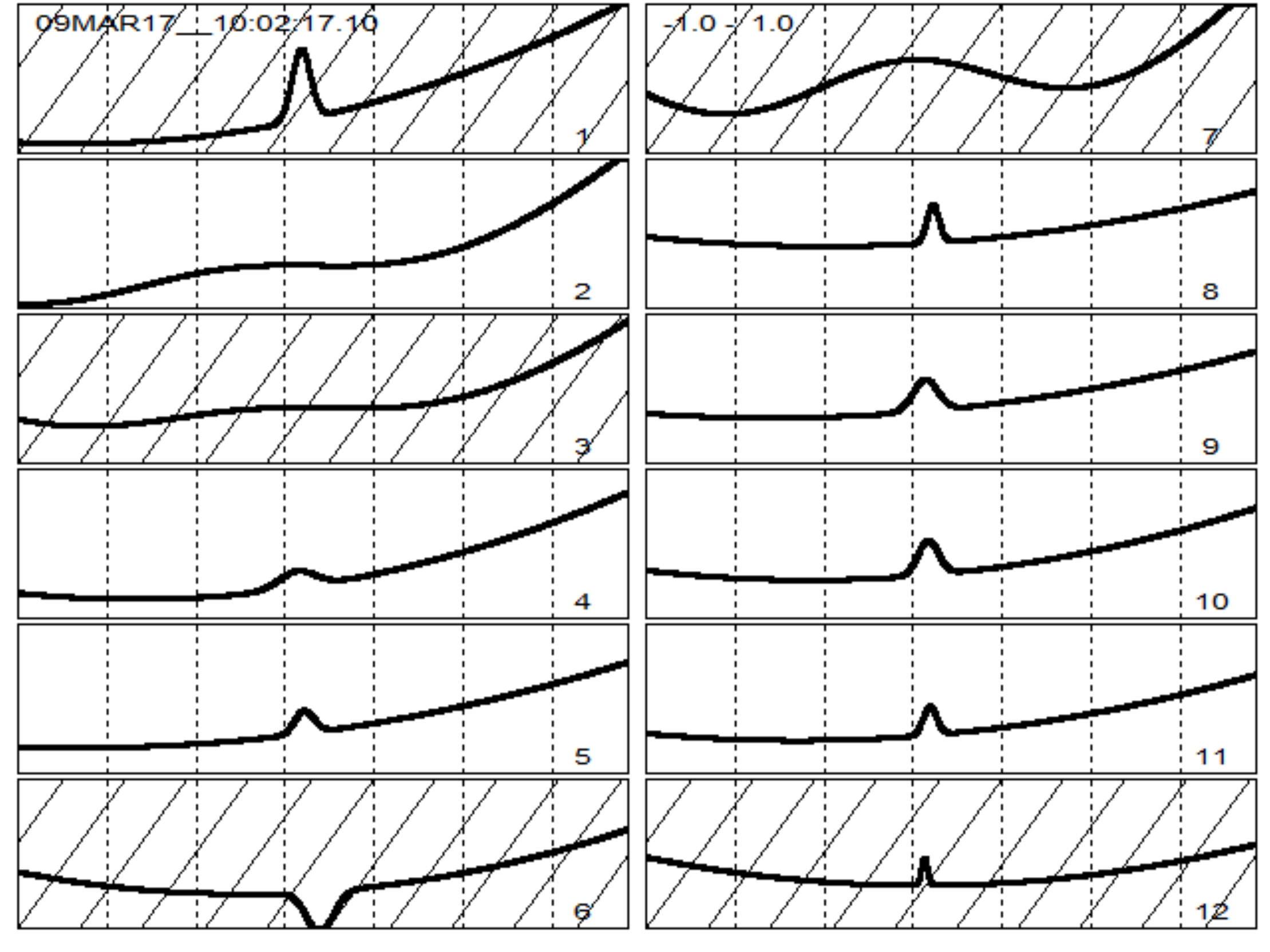}
  \caption{Fits with the variance set to 0.01 for every point.}
  \label{fig:multiyfNoWeights}
\end{figure}

In \fig{multiyfNoWeights} we set the variance at 0.01 (\degr C)$^2$ for every point. (In a least-squares fit only the {\em variation} of weights has any effect. Their absolute value merely sets the $\chi^2$ scale.) The only major effect (cf. \fig{multiyf}) is that zone 2 joins zone 3 in misbehaving. Our final results use the overall variance from the data files.

Finally, \fig{multiyfPulse} shows a fit to the data with a nearly square pulse. The rise and fall times are fixed at 1\,yr. (If they are left free, the fit runs wild. If fixed at any value small compared with the width, they have little effect.) Therefore, as before, the program retains six free parameters: three for the polynomial plus the start time, amplitude and width of the pulse. A pulse shows up around the right time in all eight eligible zones. However, we eventually settled on the Gaussian, to avoid arbitrary fixed parameters and because it seemed more like the form suggested by the data.

\subsection{Final Cut}

\subsubsection{Goodness of Fit} 
\begin{figure}[p] 
  \centering
  \includegraphics[bb=0 0 640 480,width=5.32in,height=4.0in,keepaspectratio]{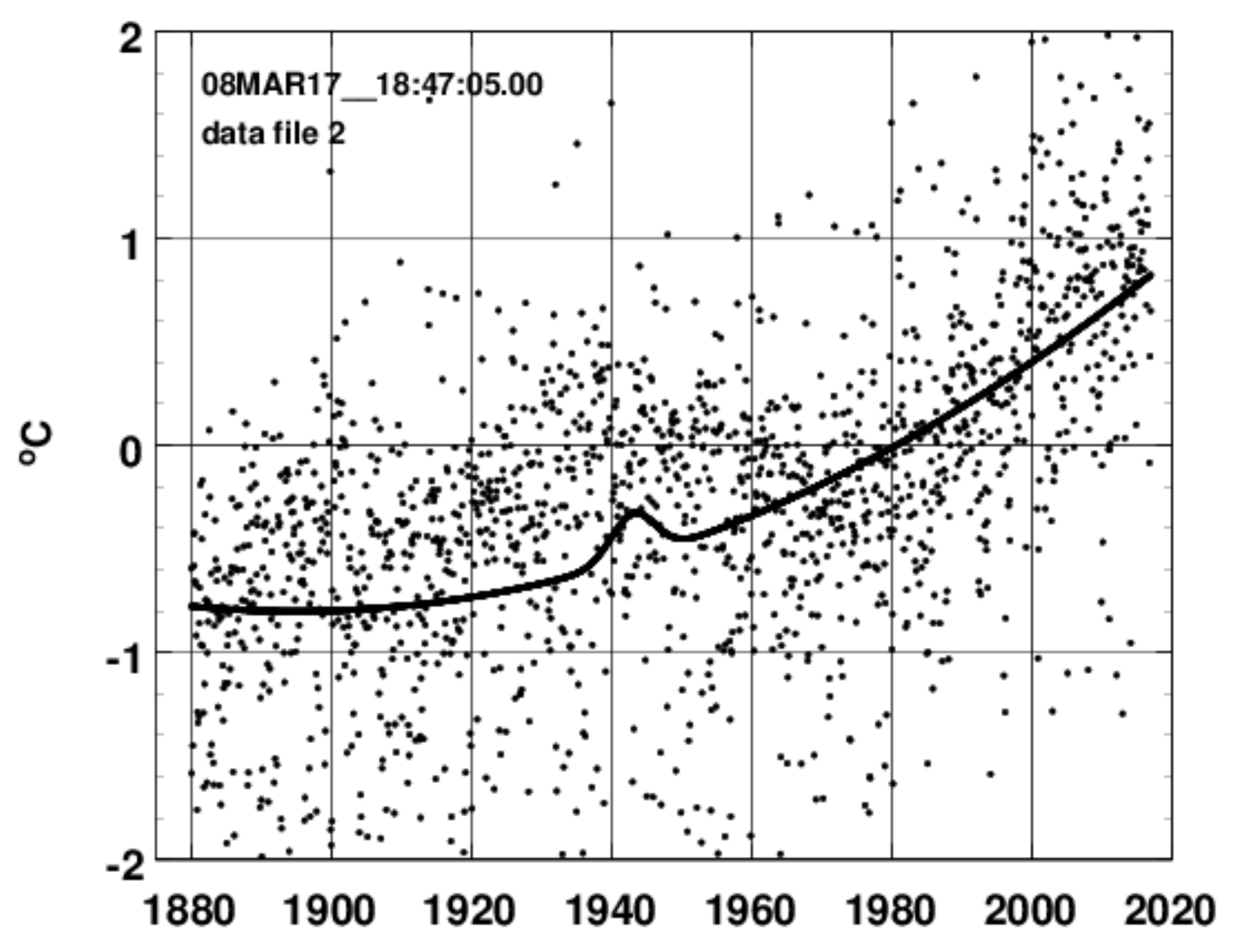}
  \caption{Worst fit used (zone 2).}
  \label{fig:fitGAU02}
\end{figure}

\begin{figure}[p] 
  \centering
  \includegraphics[bb=0 0 640 480,width=5.32in,height=4.0in,keepaspectratio]{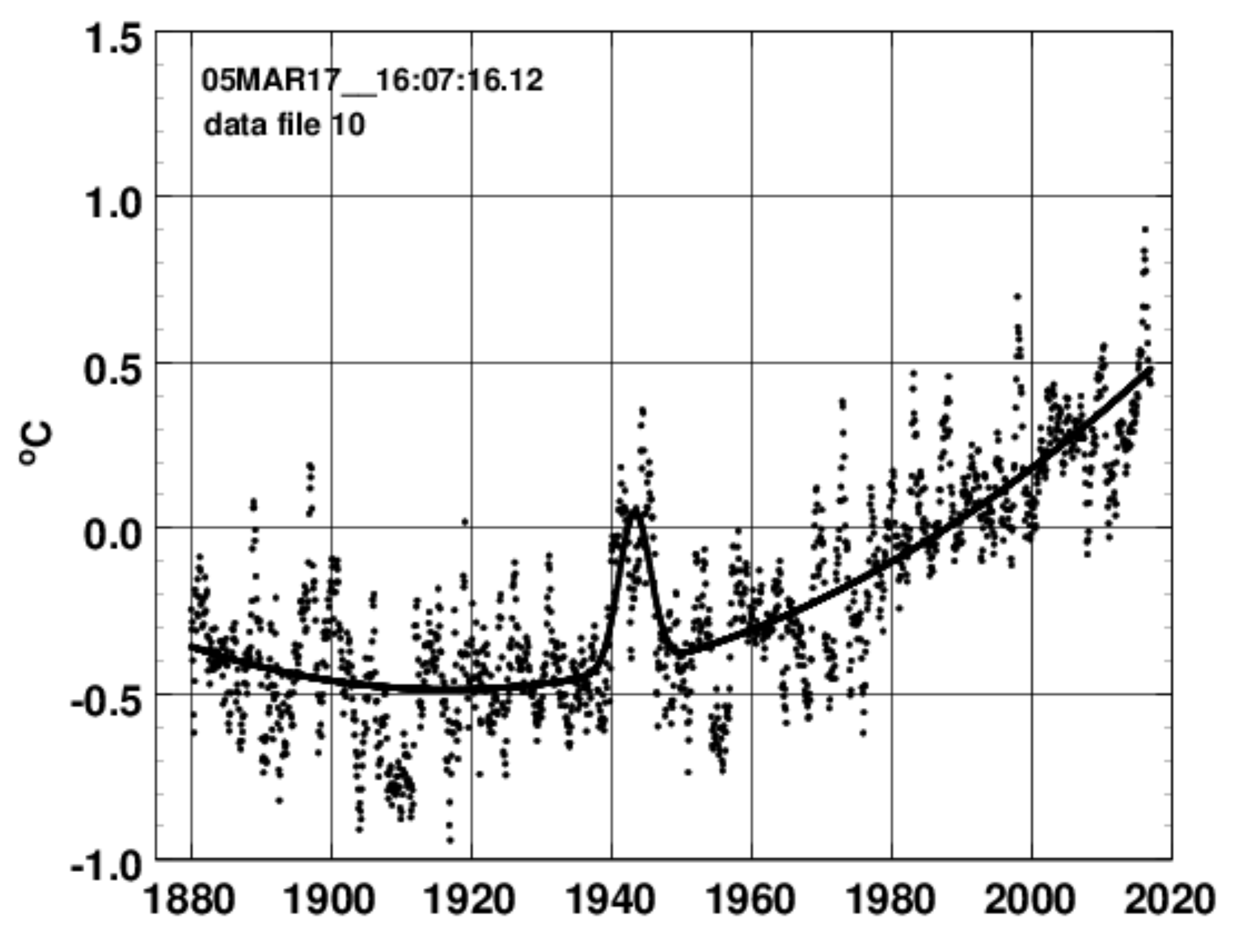}
  \caption{Best fit used (zone 10).}
  \label{fig:fitGAU10}
\end{figure}

\begin{figure}[p] 
  \centering
  \includegraphics[bb=0 0 640 480,width=5.32in,height=4.0in,keepaspectratio]{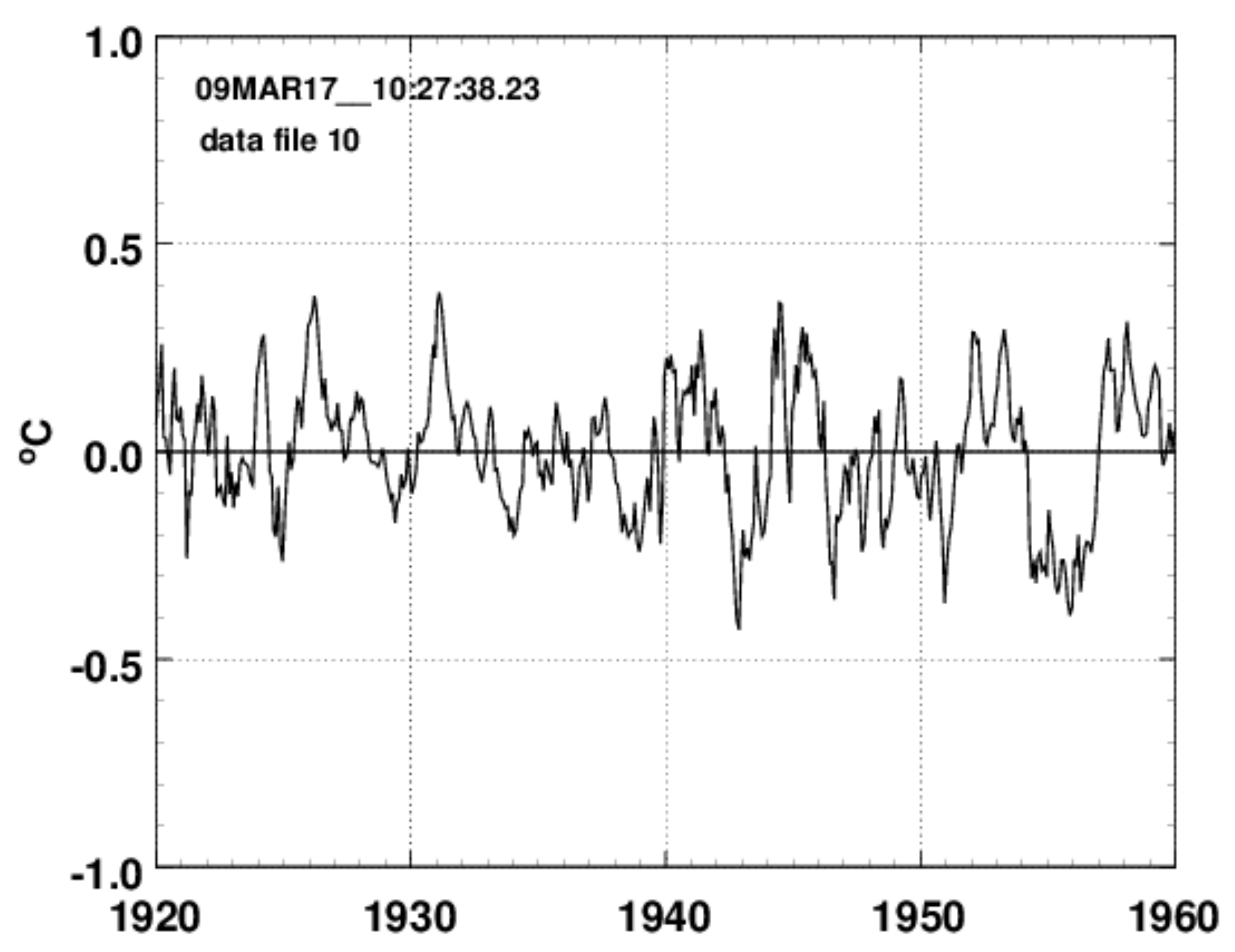}
  \caption{Residuals from best fit (zone 10) showing the El Ni\`no Southern Oscillation (ENSO) with the long-term trend and WW\caproman{2} bump removed.}
  \label{fig:resgau10}
\end{figure}

Let us now examine fit quality. Figs.\,\ref{fig:fitGAU02} and \ref{fig:fitGAU10} show the worst (zone 2) and best (zone 10) cases along with the fitted data. The scatter in zone 2 (note scale) seems rather random, with a suggestion of a denser and a broader belt. The scatter in zone 10 is entirely different and comes from ENSO, a high frequency systematic variation. \fig{resgau10} is a zoomed-in view of the residuals (data minus fit function) showing a clear $\approx5$\,yr periodicity. It is hard to say exactly what the underlying random measurement error is, but by eye its rms value appears to be 0.1\,\degr C or less.

\begin{figure}[p] 
  \centering
  \includegraphics[bb=0 0 640 480,width=5.32in,height=4.0in,keepaspectratio]{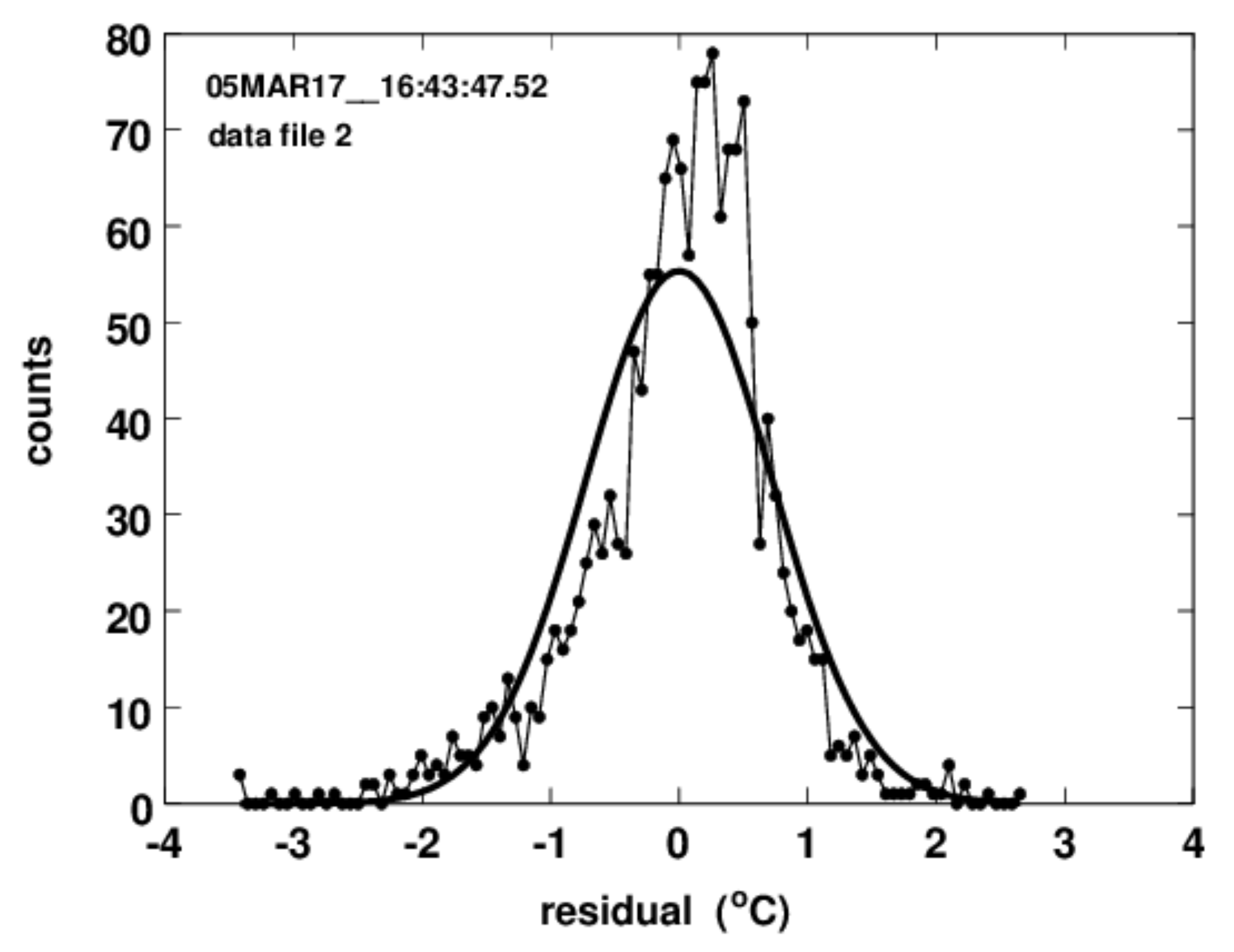}
  \caption{Frequency distribution of residuals for zone 2 (worst fit): rms = 0.726 \degr C, skewness = -0.7004, kurtosis = 1.9926\,.}
  \label{fig:freqGAU02}
\end{figure}

Goodness of fit can be quantified somewhat by looking at the frequency distribution of residuals which, for purely random errors, should be Gaussian. \fig{freqGAU02} is the distribution for zone 2. The Gaussian shown for reference is one whose mean, $\sigma$ and area equal those of the data. The rms value, skewness and kurtosis of the data are given in the caption. (The residual mean is driven to zero by the fit itself. The skewness and kurtosis of a Gaussian are zero.) In zone 2 the residuals are leptokurtic and negatively skewed. That confirms the impression from \fig{fitGAU02} of two groups of measurements, one broader and somewhat more negative than the other.

\begin{figure}[p] 
  \centering
  \includegraphics[bb=0 0 640 480,width=5.32in,height=4.0in,keepaspectratio]{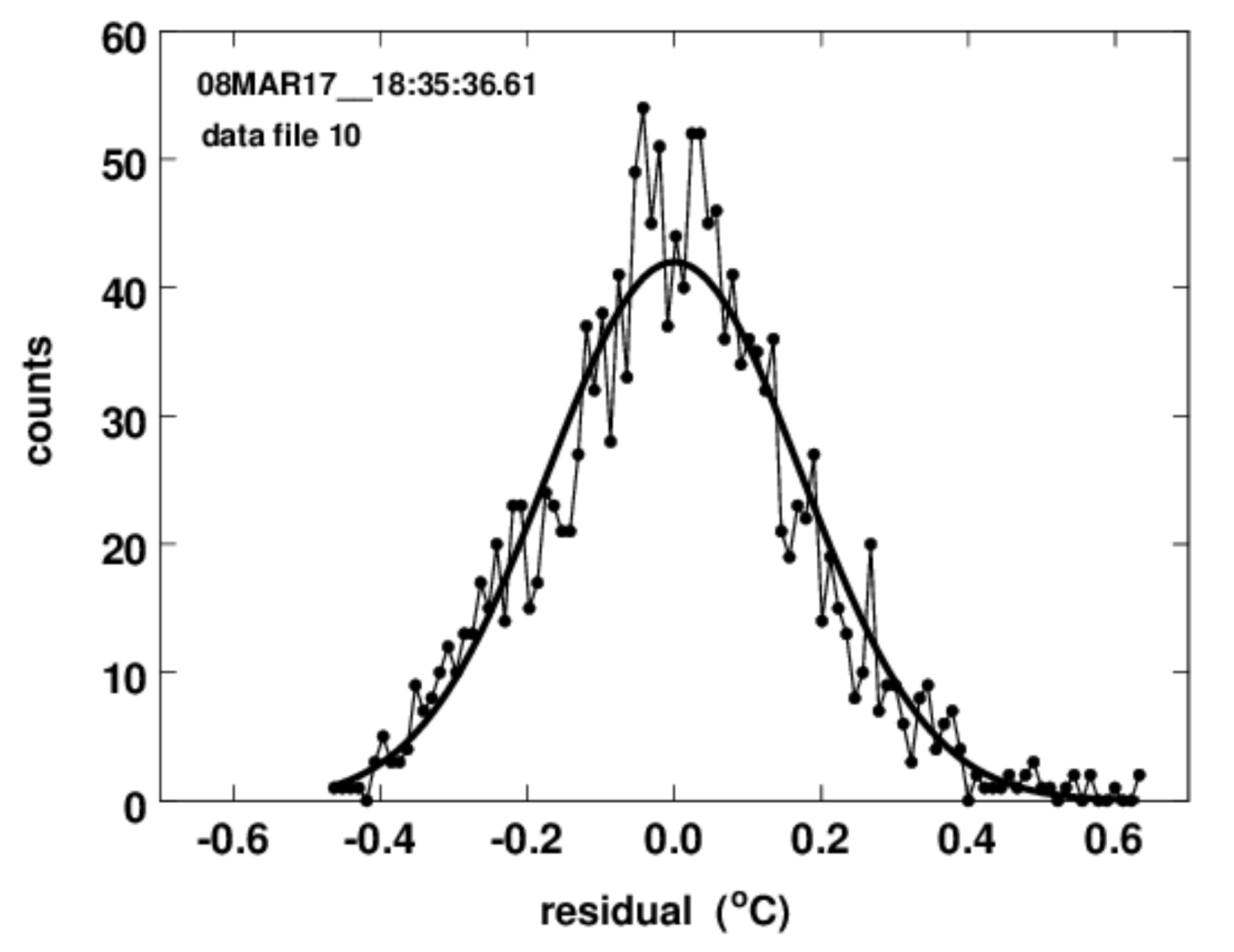}
  \caption{Frequency distribution of residuals for zone 10 (best fit): rms = 0.1728 \degr C, skewness = 0.2033, kurtosis = 0.2622\,.}
  \label{fig:freqGAU10}
\end{figure}

The corresponding \fig{freqGAU10} for zone 10 shows nearly Gaussian behavior despite the fact that the residuals are clearly not random (\fig{resgau10}). Evidently ENSO averaged over a long time is pseudo-random. The rms (0.1728\,\degr C) largely comes from ENSO rather than true measurement error.

\begin{figure}[p] 
  \centering
  \includegraphics[bb=0 0 640 480,width=5.32in,height=4.0in,keepaspectratio]{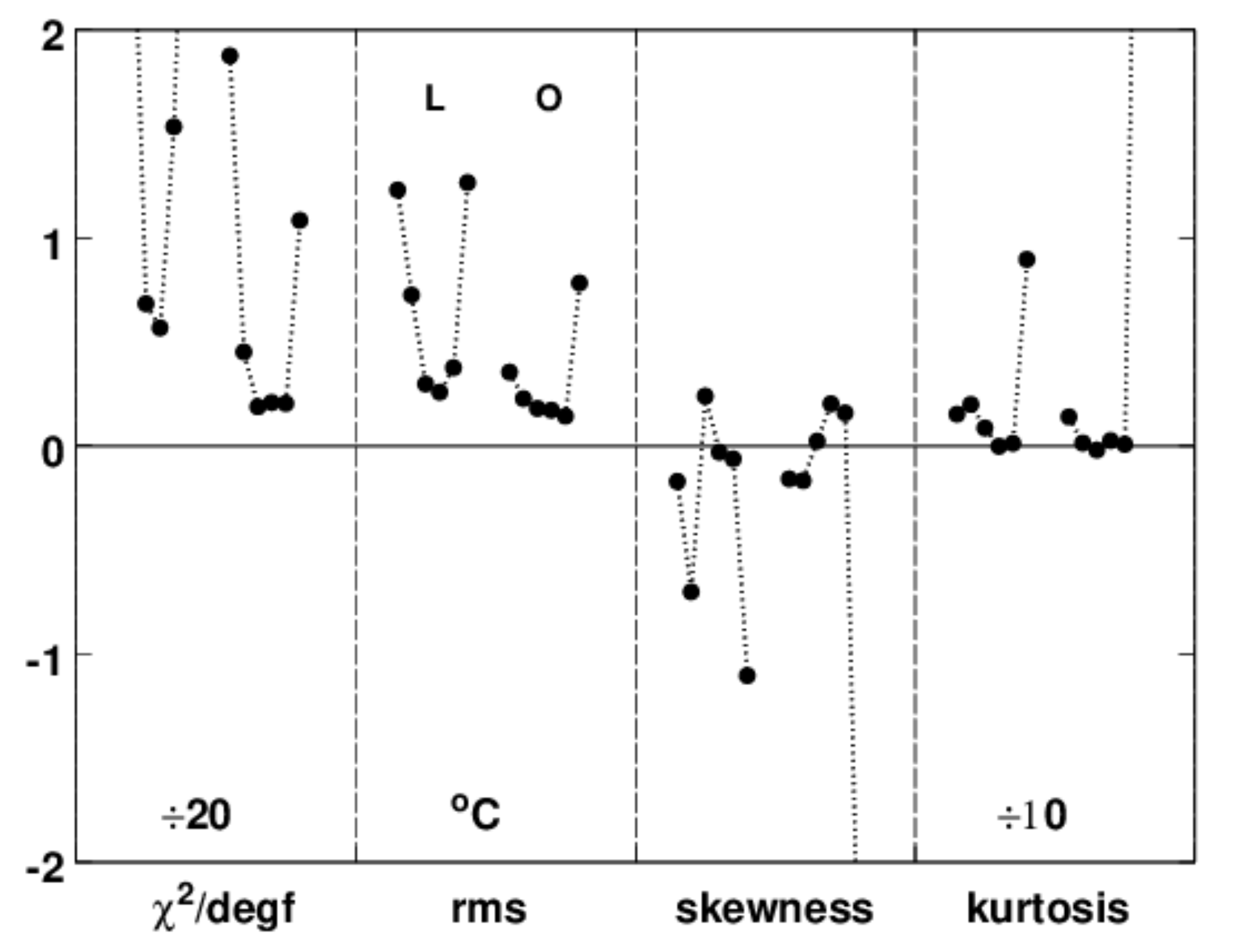}
  \caption{Summary of goodness-of-fit criteria.}
  \label{fig:goodness}
\end{figure}

Trends in goodness-of-fit are summarized by \fig{goodness} which shows, in four columns, four statistics: $\chi^2/(M-n)$ and the rms value, skewness and kurtosis of the frequency distribution of residuals (note scale factors). Two subgroups in each column show results from the land-based (L) and ocean-based (O) zones, six in each, with occasional off-scale entries.

$\chi^2/(M-n)$ improves towards the equator for both L and O (better data) although the best value attained is $\approx4$ rather than 1 for reasons already noted. The rms deviation varies in the same fashion, with O being better than L. All four measures justify our choice of zone\,2 as the worst fit (excepting the polar zones). A case could be made for zone\;9 being better than 10, but that choice was used for illustration only and does not affect the bottom line.

\subsubsection{Final Parameters: Tables}
Table\,\ref{tbl:bkgParms} lists the fitted parameters of the quadratic background in terms of transformed time $u$, \eqns{u}{P}. They are given for all twelve zones but only the seven boldfaced ones are included in the final weighted average (last row). Table\,\ref{tbl:GaussianParms}, in the same format, lists the parameters of the Gaussian bump in terms of $u$, \eqn{G}. For convenience, Tables\,\ref{tbl:bkgParmsO} and \ref{tbl:GaussianParmsO} list, similarly, parameters in terms of years, but our discussion will mainly concern Tables\,\ref{tbl:bkgParms} and \ref{tbl:GaussianParms}. Trends evident in those tables are plotted in \fig{parms} and discussed below. 

\begin{figure}[p] 
  \centering
  \includegraphics[bb=0 0 640 480,width=5.32in,height=4.0in,keepaspectratio]{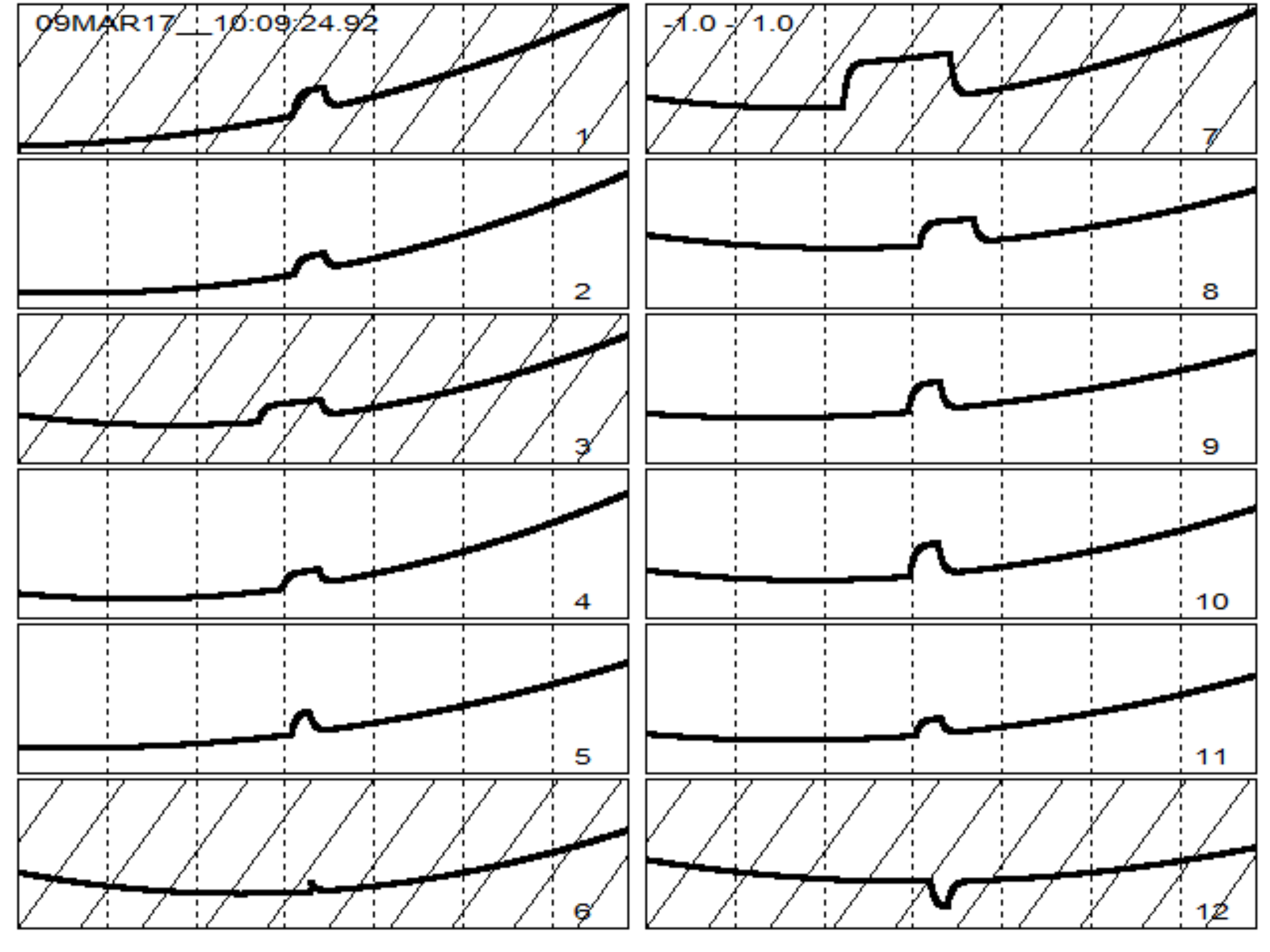}
  \caption{Fits using a nearly square pulse instead of a Gaussian. Rise and fall times are fixed at 0.015 ($\approx1$\,yr).}
  \label{fig:multiyfPulse}
\end{figure}

Table\,\ref{tbl:GaussianParmsO} has extra columns giving the {\em area} of the Gaussian and its error. Because the area depends on both amplitude and $\sigma$, its relative error is greater. By the way, the square-pulse fit (cf. \fig{multiyfPulse}) yields area\,$=1.968\pm0.435$\,\degr C\;yr in agreement with the Gaussian area\,$=1.948\pm0.577$\,\degr C\;yr. (The raw Gaussian and pulse parameters are, of course, not directly comparable.)

\subsubsection{Final Parameters: Graph}
\begin{figure}[p] 
  \centering
  \includegraphics[bb=0 0 640 480,width=5.32in,height=4.0in,keepaspectratio]{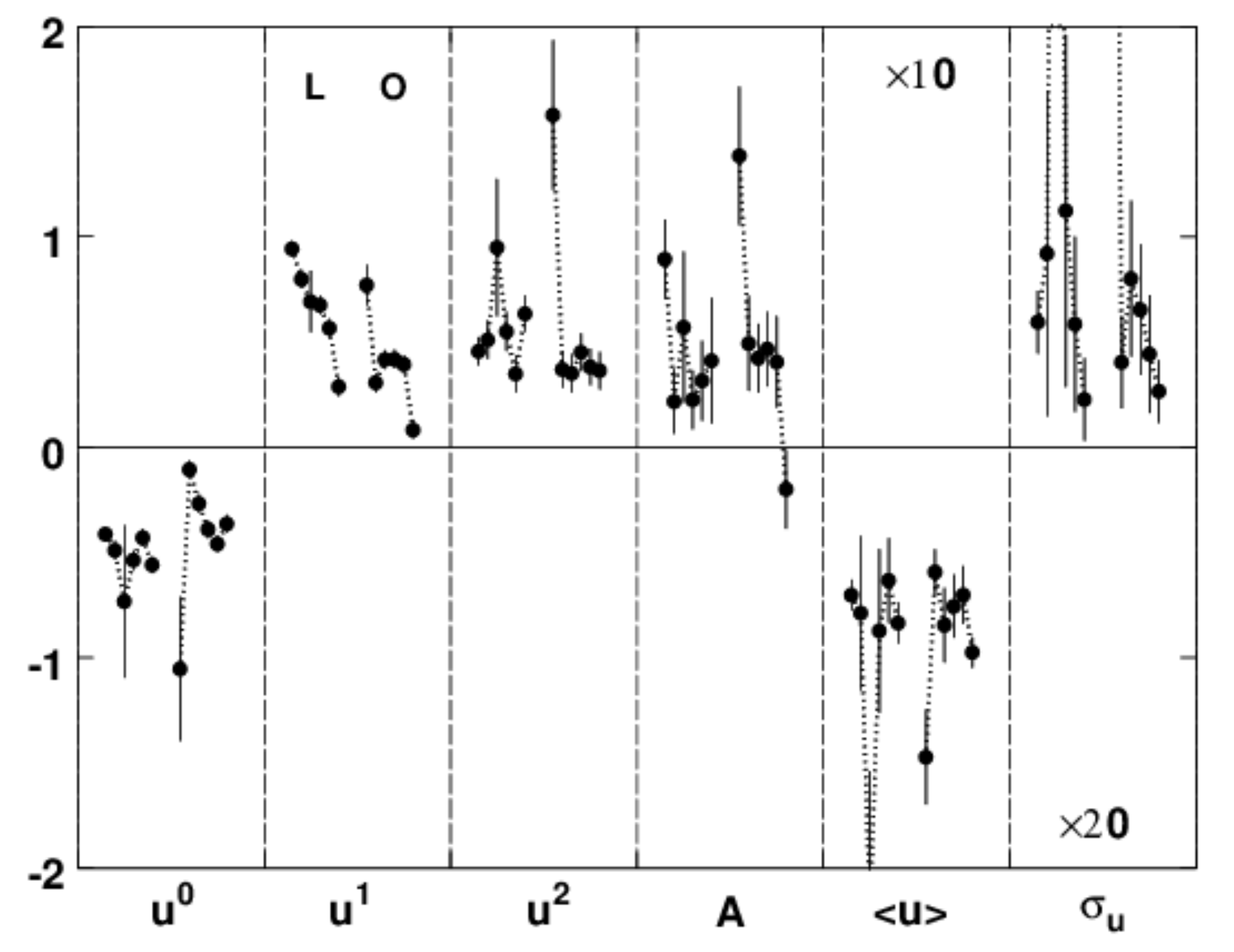}
  \caption{Summary of background and pulse parameters (Tables\,\ref{tbl:bkgParms} and \ref{tbl:GaussianParms}).}
  \label{fig:parms}
\end{figure}

Fitted parameters are plotted in \fig{parms} comprising both background and Gaussian (Tables\,\ref{tbl:bkgParms} and \ref{tbl:GaussianParms}). Error bars are $1\sigma$ (note scale factors) and all fits are shown including those dropped from the weighted average.

Starting from the left, the difference between L and O constant terms may be result from slightly different definitions of temperature `anomaly'. In any case, it has no profound consequences. More interesting are the next two blocks both showing linear and quadratic terms larger for L than for O. Both the rate and the acceleration of surface temperature increase seem to be greater for atmospheric than for ocean measurements. The disequilibrium between the atmosphere and the oceans seems to be increasing. Note that the relative errors in those numbers are small except for excluded zone\,3.

The Gaussian amplitude block confirms our earlier impression that the WW2 bump is smaller (but still present) in land-based measurements, while the last two columns show that the time of ocurrence and duration are, within errors, the same. 
  
\subsubsection{Predictions of Surface Temperature}
\begin{figure}[p] 
  \centering
  \includegraphics[bb=0 0 640 480,width=5.32in,height=4.0in,keepaspectratio]{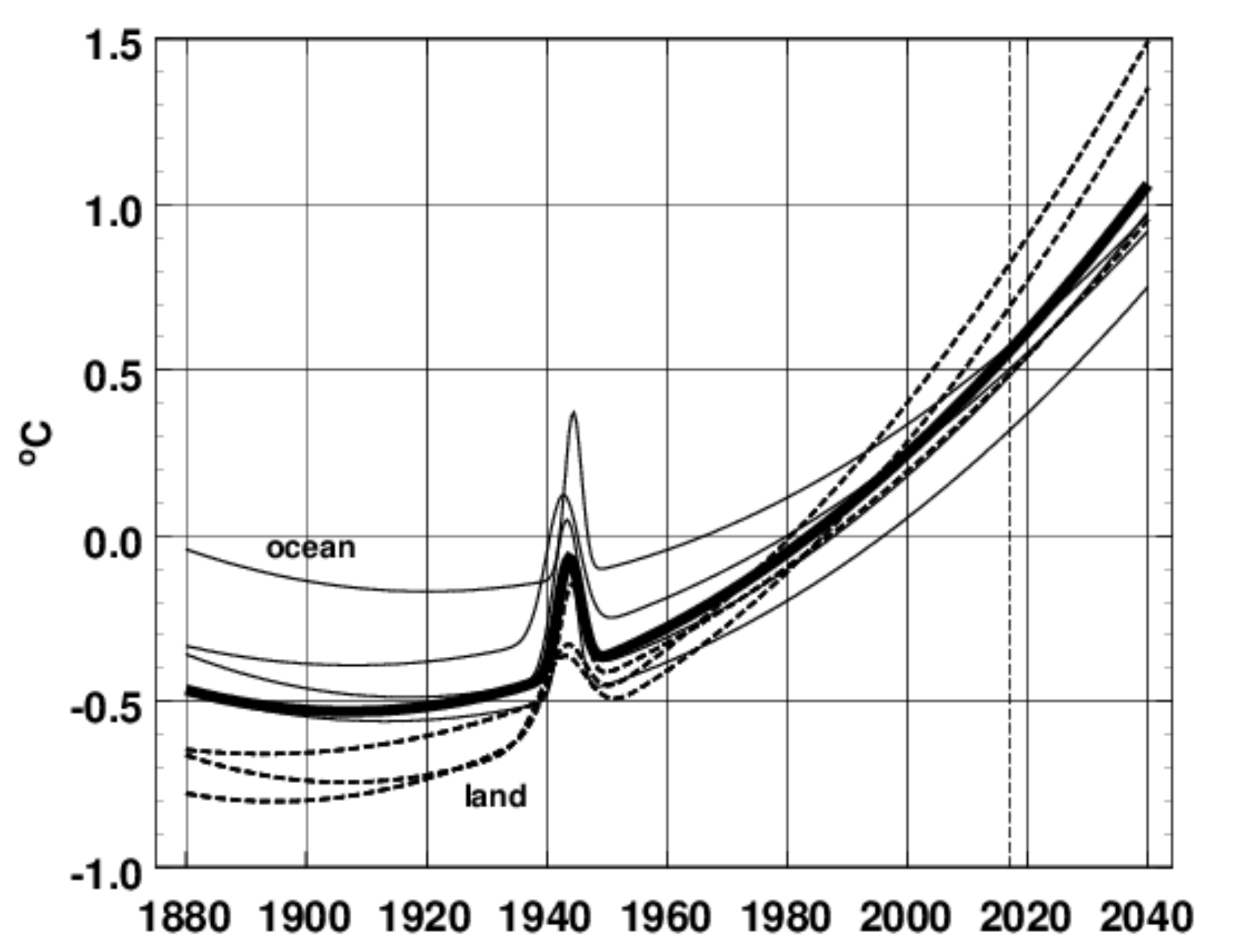}
  \caption{Fitted and predicted surface temperatures. Dashed line: land; light line: ocean; bold line: weighted average.}
  \label{fig:predicted}
\end{figure}

Although this work was motivated by and is preoccupied with the WW2 bump, the long term trends which emerge as a byproduct are far more consequential. \fig{predicted} shows our fits extended by two decades. The differences between atmospheric (dashed) and ocean (line) trends already noted are seen in full, and the bold line reflects the combination (last line of Tables\,\ref{tbl:bkgParms} and \ref{tbl:GaussianParms}). It predicts an 0.5\,\degr C rise in surface temperature over the next 20\,years if nothing changes.

\begin{figure}[p] 
  \centering
  \includegraphics[bb=0 0 640 480,width=5.32in,height=4.0in,keepaspectratio]{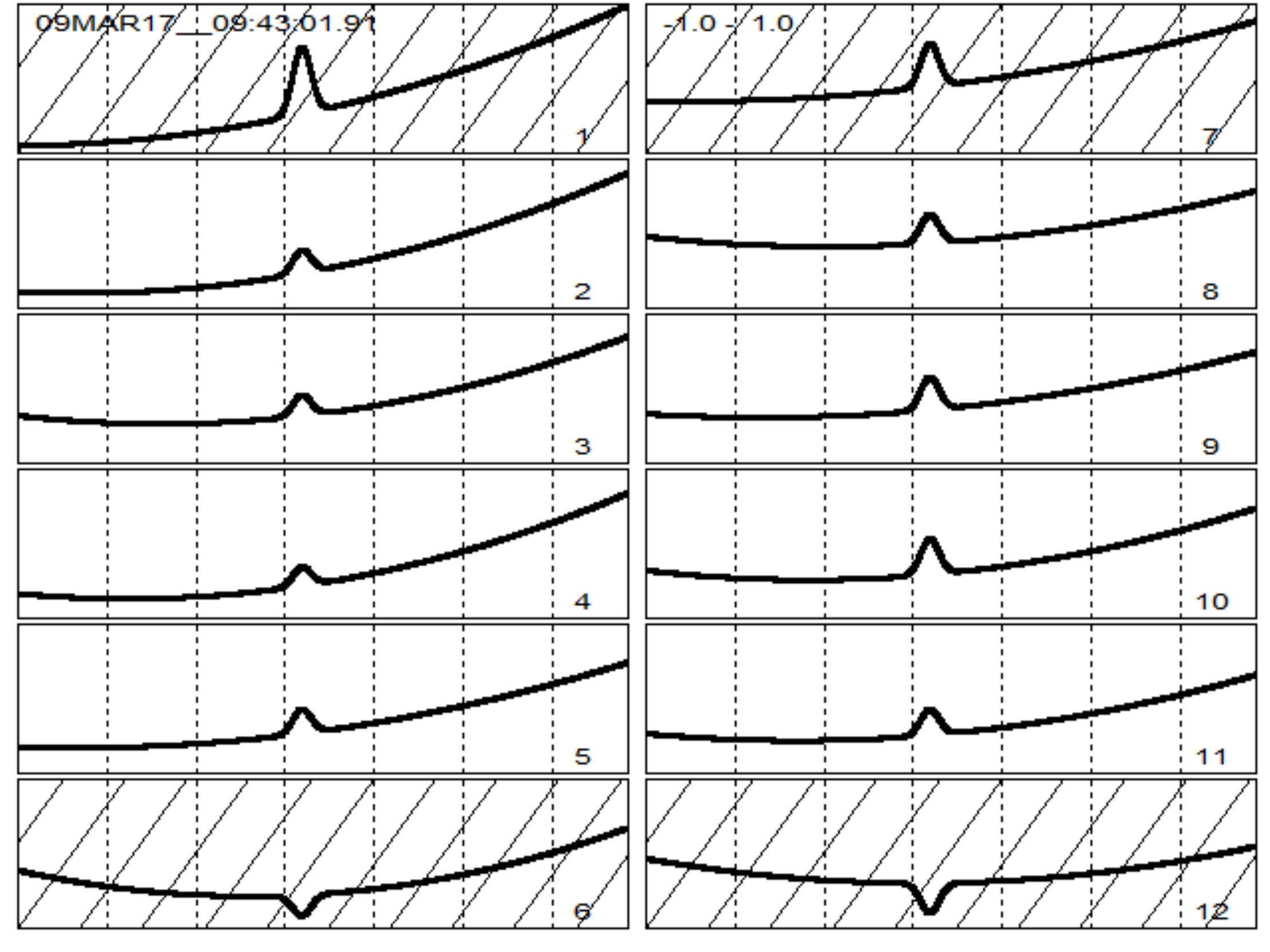}
  \caption{Fits with fixed Gaussian mean (-0.07) and width (0.03). The Gaussian amplitude is adjustable by the program.}
  \label{fig:multiyflock}
\end{figure}

\begin{figure}[p] 
  \centering
  \includegraphics[bb=0 0 640 480,width=5.32in,height=4.0in,keepaspectratio]{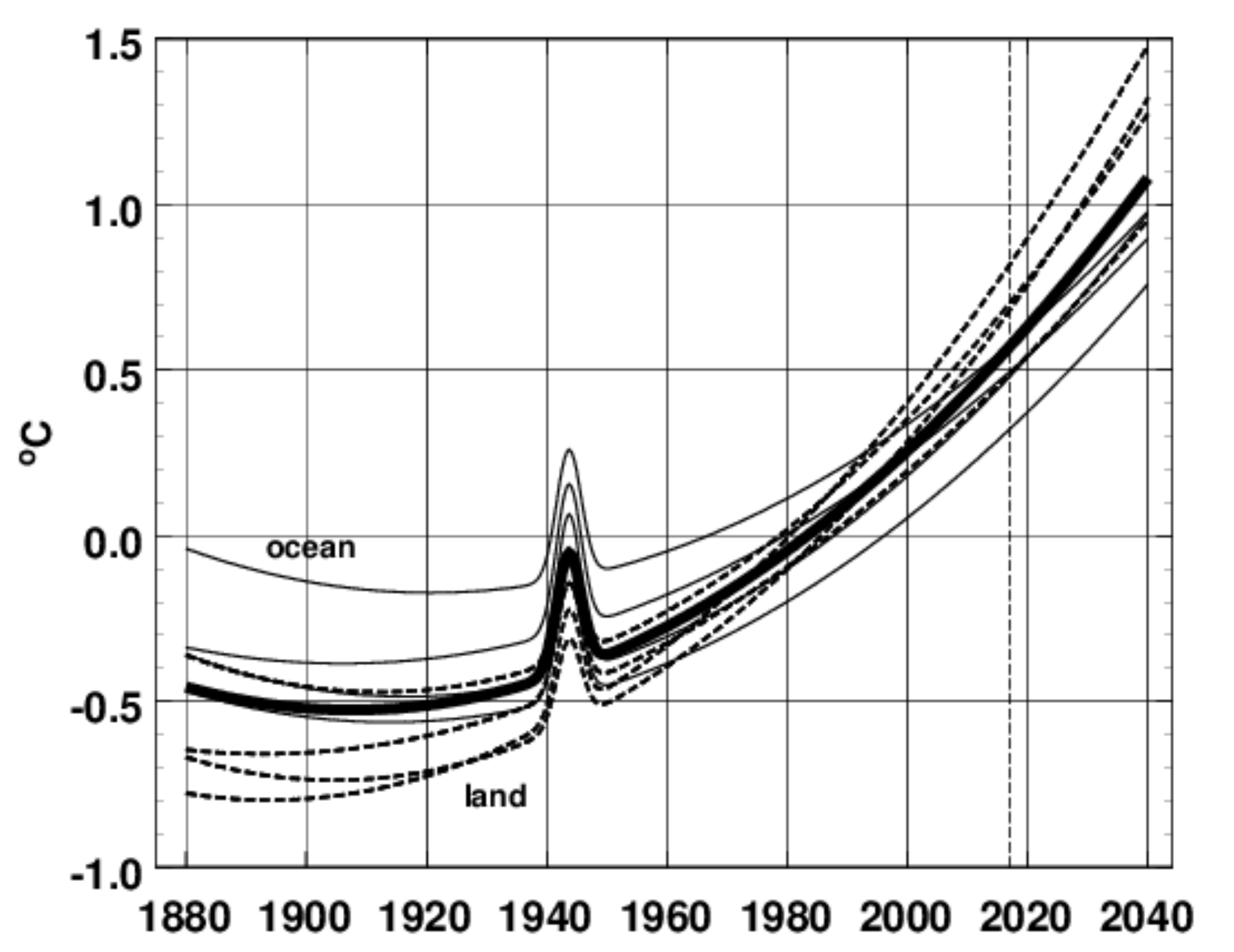}
  \caption{The same as Fig.\,\ref{fig:predicted}, but Gaussian position and width locked cf. Fig.\,\ref{fig:multiyflock} and all equatorial and median zones accepted. The vertical dashed line is at 2017.}
  \label{fig:PredLock}
\end{figure}

A cleaner but essentially similar picture (\fig{PredLock}) is obtained by holding the Gaussian position and width fixed and using all eight non-polar zones (cf. \fig{multiyflock}).

\section{Discussion and Summary}
Using parametric analysis (curve fitting) we have extracted numbers, with errors, that encapsulate the overall trend of atmospheric and ocean surface temperature including a bump around WW2. The bump is evident in many previous graphs but seems to have attracted little comment. It is a robust feature of the data, appearing in both the atmospheric and ocean signal. Its amplitude differs from 0 by 5.2\,$\sigma$ (cf. Table\,\ref{tbl:GaussianParms}) so the hypothesis of a statistical fluctuation can be ruled out. Three explanations would seem to remain:
\begin{enumerate}
\renewcommand{\theenumi}{\Alph{enumi}}
\item A systematic measuring error came and went around WW2.
\item Coincidence: an unrelated positive climatic event occurred around WW2.
\item The bump is a consequence of human activity during WW2.
\end{enumerate}

(A) is superficially plausible, WW2 being a time of global upheaval, until one recalls that the aggregation of measurements analyzed here began well after the war. There is no reason to think that land stations changed their protocols at the start of the war and changed them back at the end. Nor did the global distribution of stations (which might use different protocols) change very much during the war \cite{stationData}. The most likely systematic candidate is the well-known `bucket' effect \cite{Thompson2008}. Against that it can be argued first, that it was unipolar, more or less coincident with the end of WW2\,; second, that it affected only ocean measurements and third, that ERSST.v4 attempts to correct it \cite{Huang2015}. 

(B) can never be ruled out, logically speaking, but theoretical or phenomenological arguments why an event should have occurred at just such a time should be advanced to support it.

(C) seems the simplest and most likely. Indeed, if one believes that human activity affects global temperature, one cannot dismiss out-of-hand the idea that WW2, an orgy of combustion of various fuels, had an effect. Note that we are speaking of combustion relative to that era, not the present day. It would be illuminating to quantify the use of fossil fuels (and other war-related burning) relative to adjacent time periods, but that is a job for the historian. Such data might enable one to test climate models. 

If one accepts (C) it is encouraging that temperature recovered to baseline rather quickly. In that era at least, temperature seems to have been stable with respect to small increases in greenhouse gases. There is no guarantee that is still so, but it gives grounds for hope.

As to the background global temperature increase, our extrapolation holds independent of one's opinion of the WW2 bump, and there is little more to say. Our linear term, $0.747\pm0.024$\;\degr C/century (Table\,\ref{tbl:bkgParmsO}) agrees with $0.735\pm0.068$\;\degr C/century from an independent fit to ERSST.v4 cited by Huang et al. \cite{Huang2015}.\footnote{~We took the weighted average of the first column of their Table\,2.} If we omit the Gaussian and quadratic term in our program and just fit everything with a straight line, the weighted average of the usual 7 zones is $0.742\pm0.024$\;\degr C/century agreeing with the fuller analysis. The Gaussian and quadratic terms seem to have negligible effect on the slope.

\section{Acknowledgements}
We thank Harvard University, the Physics Department, and the Laboratory for Particle Physics and Cosmology for generous support. Needless to say, any opinions lurking in this report, which we have tried to make as objective as possible, are our own and not those of the University.

Any study of this kind depends on unnamed persons, in their thousands, who took the data, and to the scientists and staff at NOAA and elsewhere who collected, refined and published it.


\clearpage
\listoftables
\listoffigures

\clearpage
\begin{table}[h]
\setlength{\tabcolsep}{10pt}
\begin{center}
\begin{tabular}{rrrrrrr}
\multicolumn{1}{c}{data}&           
\multicolumn{1}{c}{$p_1$}&           
\multicolumn{1}{c}{$\sigma_{p_1}$}&           
\multicolumn{1}{c}{$p_2$}&           
\multicolumn{1}{c}{$\sigma_{p_2}$}&           
\multicolumn{1}{c}{$p_3$}&           
\multicolumn{1}{c}{$\sigma_{p_3}$}\\           
\noalign{\vspace{2pt}}
\multicolumn{1}{c}{file}&           
\multicolumn{1}{c}{\degr C/$u^0$}&           
\multicolumn{1}{c}{\degr C/$u^0$}&           
\multicolumn{1}{c}{\degr C/$u^1$}&           
\multicolumn{1}{c}{\degr C/$u^1$}&           
\multicolumn{1}{c}{\degr C/$u^2$}&           
\multicolumn{1}{c}{\degr C/$u^2$}\\           
\noalign{\vspace{6pt}}
       1&    -0.4124&     0.0291&     0.9443&     0.0336&     0.4567&     0.0650\\
\bf    2&\bf -0.4890&\bf  0.0433&\bf  0.7997&\bf  0.0428&\bf  0.5110&\bf  0.0904\\
       3&    -0.7330&     0.3614&     0.6930&     0.1446&     0.9493&     0.3247\\
\bf    4&\bf -0.5363&\bf  0.0450&\bf  0.6766&\bf  0.0428&\bf  0.5508&\bf  0.0926\\
\bf    5&\bf -0.4288&\bf  0.0407&\bf  0.5665&\bf  0.0427&\bf  0.3497&\bf  0.0873\\
       6&    -0.5585&     0.0383&     0.2880&     0.0427&     0.6351&     0.0843\\
       7&    -1.0532&     0.3402&     0.7709&     0.0952&     1.5793&     0.3550\\
\bf    8&\bf -0.1046&\bf  0.0395&\bf  0.3070&\bf  0.0427&\bf  0.3708&\bf  0.0857\\
\bf    9&\bf -0.2670&\bf  0.0423&\bf  0.4175&\bf  0.0428&\bf  0.3521&\bf  0.0892\\
\bf   10&\bf -0.3898&\bf  0.0413&\bf  0.4191&\bf  0.0427&\bf  0.4514&\bf  0.0880\\
\bf   11&\bf -0.4582&\bf  0.0397&\bf  0.3938&\bf  0.0427&\bf  0.3824&\bf  0.0860\\
      12&    -0.3630&     0.0408&     0.0831&     0.0427&     0.3644&     0.0874\\
\noalign{\vspace{6pt}}
\bf avg&\bf -0.3752&\bf  0.0157&\bf  0.5112&\bf  0.0162&\bf  0.4210&\bf  0.0334\\
\end{tabular}
\caption{Background parameters and errors. The last line is the weighted average of bold-faced entries. $u$ is a dimensionless time variable running from -1 to 1 as years run from 1880 through 2016.}\label{tbl:bkgParms}
\end{center}
\end{table}

\begin{table}[h]
\setlength{\tabcolsep}{10pt}
\begin{center}
\begin{tabular}{rrrrrrr}
\multicolumn{1}{c}{data}&           
\multicolumn{1}{c}{$p_6$}&           
\multicolumn{1}{c}{$\sigma_{p_6}$}&           
\multicolumn{1}{c}{$p_7$}&           
\multicolumn{1}{c}{$\sigma_{p_7}$}&           
\multicolumn{1}{c}{$p_8$}&           
\multicolumn{1}{c}{$\sigma_{p_8}$}\\           
\noalign{\vspace{2pt}}
\multicolumn{1}{c}{file}&           
\multicolumn{1}{c}{\degr C}&           
\multicolumn{1}{c}{\degr C}&           
\multicolumn{1}{c}{$u$}&           
\multicolumn{1}{c}{$u$}&           
\multicolumn{1}{c}{$u$}&           
\multicolumn{1}{c}{$u$}\\           
\noalign{\vspace{6pt}}
       1&     0.8950&     0.1869&    -0.0702&     0.0071&     0.0297&     0.0074\\
\bf    2&\bf  0.2184&\bf  0.1519&\bf -0.0788&\bf  0.0364&\bf  0.0461&\bf  0.0386\\
       3&     0.5709&     0.3572&    -0.2257&     0.0716&     0.3754&     0.1640\\
\bf    4&\bf  0.2265&\bf  0.1386&\bf -0.0872&\bf  0.0388&\bf  0.0562&\bf  0.0418\\
\bf    5&\bf  0.3162&\bf  0.1884&\bf -0.0633&\bf  0.0201&\bf  0.0293&\bf  0.0208\\
       6&     0.4114&     0.2970&    -0.0835&     0.0096&     0.0113&     0.0097\\
       7&     1.3856&     0.3280&    -0.1473&     0.0224&     0.3389&     0.0564\\
\bf    8&\bf  0.4939&\bf  0.2248&\bf -0.0592&\bf  0.0107&\bf  0.0202&\bf  0.0109\\
\bf    9&\bf  0.4230&\bf  0.1622&\bf -0.0845&\bf  0.0176&\bf  0.0401&\bf  0.0184\\
\bf   10&\bf  0.4669&\bf  0.1759&\bf -0.0756&\bf  0.0148&\bf  0.0327&\bf  0.0155\\
\bf   11&\bf  0.4058&\bf  0.2153&\bf -0.0702&\bf  0.0136&\bf  0.0221&\bf  0.0140\\
      12&    -0.1995&     0.1861&    -0.0975&     0.0071&     0.0133&     0.0074\\
\noalign{\vspace{6pt}}
\bf avg&\bf  0.3394&\bf  0.0652&\bf -0.0692&\bf  0.0062&\bf  0.0276&\bf  0.0064\\
\end{tabular}
\caption{Gaussian parameters and errors.  The last line is the weighted average of bold-faced entries. $u$ is a dimensionless time variable running from -1 to 1 as years run from 1880 through 2016. $p_6$, $p_7$, $p_8$ are, respectively, the Gaussian amplitude, mean and rms deviation.}\label{tbl:GaussianParms}
\end{center}
\end{table}

\clearpage
\begin{table}[h]
\setlength{\tabcolsep}{10pt}
\begin{center}
\begin{tabular}{rrrrrrr}
\multicolumn{1}{c}{file}&           
\multicolumn{1}{c}{$p'_1$}&           
\multicolumn{1}{c}{$\sigma_{p'_1}$}&           
\multicolumn{1}{c}{$p'_2$}&           
\multicolumn{1}{c}{$\sigma_{p'_2}$}&           
\multicolumn{1}{c}{$p'_3$}&           
\multicolumn{1}{c}{$\sigma_{p'_3}$}\\           
\noalign{\vspace{2pt}}
\multicolumn{1}{c}{}&           
\multicolumn{1}{c}{\degr C}&           
\multicolumn{1}{c}{\degr C}&           
\multicolumn{1}{c}{\degr C/hyr}&           
\multicolumn{1}{c}{\degr C/hyr}&           
\multicolumn{1}{c}{\degr C/hyr$^2$}&           
\multicolumn{1}{c}{\degr C/hyr$^2$}\\           
\noalign{\vspace{6pt}}
       1&    -0.4124&     0.0291&     1.3794&     0.0491&     0.9745&     0.1388\\
\bf    2&\bf -0.4890&\bf  0.0433&\bf  1.1682&\bf  0.0625&\bf  1.0903&\bf  0.1930\\
       3&    -0.7330&     0.3614&     1.0123&     0.2113&     2.0257&     0.6928\\
\bf    4&\bf -0.5363&\bf  0.0450&\bf  0.9883&\bf  0.0626&\bf  1.1752&\bf  0.1975\\
\bf    5&\bf -0.4288&\bf  0.0407&\bf  0.8275&\bf  0.0624&\bf  0.7461&\bf  0.1862\\
       6&    -0.5585&     0.0383&     0.4207&     0.0624&     1.3553&     0.1798\\
       7&    -1.0532&     0.3402&     1.1260&     0.1390&     3.3698&     0.7574\\
\bf    8&\bf -0.1046&\bf  0.0395&\bf  0.4484&\bf  0.0624&\bf  0.7912&\bf  0.1829\\
\bf    9&\bf -0.2670&\bf  0.0423&\bf  0.6098&\bf  0.0625&\bf  0.7514&\bf  0.1903\\
\bf   10&\bf -0.3898&\bf  0.0413&\bf  0.6122&\bf  0.0624&\bf  0.9633&\bf  0.1878\\
\bf   11&\bf -0.4582&\bf  0.0397&\bf  0.5753&\bf  0.0624&\bf  0.8159&\bf  0.1835\\
      12&    -0.3630&     0.0408&     0.1214&     0.0624&     0.7776&     0.1864\\
\noalign{\vspace{6pt}}
\bf avg&\bf -0.3752&\bf  0.0157&\bf  0.7468&\bf  0.0236&\bf  0.8983&\bf  0.0713\\
\end{tabular}
\caption{Background parameters and errors in ordinary units (`hyr' = century). The last line is the weighted average of bold-faced rows.  }\label{tbl:bkgParmsO}
\end{center}
\end{table}

\begin{table}[h]
\setlength{\tabcolsep}{10pt}
\begin{center}
\begin{tabular}{rrrrrrrrr}
\multicolumn{1}{c}{data}&           
\multicolumn{1}{c}{$p'_6$}&           
\multicolumn{1}{c}{$\sigma_{p'_6}$}&           
\multicolumn{1}{c}{$p'_7$}&           
\multicolumn{1}{c}{$\sigma_{p'_7}$}&           
\multicolumn{1}{c}{$p'_8$}&           
\multicolumn{1}{c}{$\sigma_{p'_8}$}&           
\multicolumn{1}{c}{area}&           
\multicolumn{1}{c}{$\sigma_\mathrm{area}$}\\           
\noalign{\vspace{2pt}}
\multicolumn{1}{c}{file}&           
\multicolumn{1}{c}{\degr C}&           
\multicolumn{1}{c}{\degr C}&           
\multicolumn{1}{c}{yr}&           
\multicolumn{1}{c}{yr}&           
\multicolumn{1}{c}{yr}&           
\multicolumn{1}{c}{yr}&           
\multicolumn{1}{c}{\degr C\;yr}&           
\multicolumn{1}{c}{\degr C\;yr}\\           
\noalign{\vspace{6pt}}
       1&      0.895&      0.187&     1943.70&     0.49&      2.035&      0.507&      4.566&      1.484\\
\bf    2&\bf   0.218&\bf   0.152&\bf  1943.11&\bf  2.50&\bf   3.153&\bf   2.644&\bf   1.726&\bf   1.880\\
       3&      0.571&      0.357&     1933.05&     4.90&     25.696&     11.231&     36.775&     28.065\\
\bf    4&\bf   0.226&\bf   0.139&\bf  1942.53&\bf  2.66&\bf   3.848&\bf   2.860&\bf   2.185&\bf   2.103\\
\bf    5&\bf   0.316&\bf   0.188&\bf  1944.17&\bf  1.37&\bf   2.003&\bf   1.422&\bf   1.588&\bf   1.471\\
       6&      0.411&      0.297&     1942.78&     0.66&      0.776&      0.667&      0.800&      0.899\\
       7&      1.386&      0.328&     1938.42&     1.54&     23.197&      3.861&     80.566&     23.316\\
\bf    8&\bf   0.494&\bf   0.225&\bf  1944.45&\bf  0.73&\bf   1.385&\bf   0.749&\bf   1.715&\bf   1.212\\
\bf    9&\bf   0.423&\bf   0.162&\bf  1942.71&\bf  1.20&\bf   2.744&\bf   1.263&\bf   2.910&\bf   1.743\\
\bf   10&\bf   0.467&\bf   0.176&\bf  1943.33&\bf  1.02&\bf   2.239&\bf   1.058&\bf   2.620&\bf   1.584\\
\bf   11&\bf   0.406&\bf   0.215&\bf  1943.69&\bf  0.93&\bf   1.516&\bf   0.956&\bf   1.542&\bf   1.270\\
      12&     -0.199&      0.186&     1941.83&     0.49&      0.909&      0.506&     -0.454&     -0.494\\
\noalign{\vspace{6pt}}
\bf avg&\bf   0.339&\bf   0.065&\bf  1943.76&\bf  0.43&\bf   1.893&\bf   0.440&\bf   1.948&\bf   0.577\\
\end{tabular}
\caption{Gaussian parameters and errors in ordinary units.  The last line is the weighted average of bold-faced entries. $p'_6$, $p'_7$, $p'_8$ are, respectively, the Gaussian amplitude, mean and rms deviation. The last two columns list the Gaussian area and its $1\sigma$ error as computed from the amplitude $p_6$ and rms width $p_8$ and their errors.}\label{tbl:GaussianParmsO}
\end{center}
\end{table}

\clearpage

\clearpage

\clearpage

\end{document}